\documentclass[prl,twocolumn]{revtex4}

\usepackage{graphicx}
\usepackage [latin1]{inputenc}
\usepackage{enumerate}

\begin{document}

\title{Magnetic State Modification Induced by Superconducting Response in Ferromagnet/Superconductor Hybrids}

\author{C. Monton}

\author{F. de la Cruz}

\author{J. Guimpel}

\affiliation{Centro Atómico Bariloche \& Instituto Balseiro,\\
Comisión Nacional de Energía Atómica \& Universidad Nacional de
Cuyo, (8400) S.C. de Bariloche, Argentina}


\begin{abstract}
Magnetization measurements in superconductor/ferromagnet $Nb/Co$ superlattices show a complex behavior as a function of
temperature, applied field and sample history. In base to a simple model it is shown that this behavior is due to an interplay
between the superconductor magnetization temperature dependence, the ferromagnet magnetization time dependence, and the stray
fields of both materials. It is also shown that the magnetic state of the $Co$ layers is modified by the $Nb$ superconducting
response, implying that the problem of a superconductor/ferromagnetic heterogeneous sample has to be solved in a self-consistent
manner.
\end{abstract}



Submitted to Phys. Rev. Lett. July 12$^{th}$ 2007

\maketitle

The interaction between a superconductor, SC, and a ferromagnet, FM, in close contact at an interface, as in a superlattice, has
attracted attention in the last years due to the possibility of fabricating SC/FM hybrid devices.\cite{SpinSwitch} These
engineered materials originate the appearance of interesting physical phenomena due to the different scales and mechanisms of
interaction, like SC pair breaking effects related to exchange interaction at the interface,\cite{SpinSwitch} or electromagnetic
interaction with the stray fields of the FM both at the mesoscopic and macroscopic level.\cite{Outplane, Griegos}

Most of the research has focused on the ways in which the FM affects the SC response. For example, in the Domain Wall
Superconductivity effect, observed in SC/FM bi-layers,\cite{Outplane, Griegos}  superconductivity nucleates in those places where
the perpendicular component of the inhomogeneous FM domain structure´s stray field is close to zero. In the Spin Switch effect,
observed in FM/SC/FM trilayers,\cite{SpinSwitch} the Cooper pair, due to its finite size, experiments different average values of
the exchange field when the FM layers are ferro- or antiferro-magnetically oriented. As a consequence, the SC order parameter is
more depressed when the FM layers are ferro-magnetically oriented. This allows the control of the SC order parameter value
through an external macroscopic parameter.

In contrast, very little work has been done in exploring in which way the SC affects the magnetic state of the FM
layer.\cite{SCenFM} Recently,\cite{NuestroPRB} we have shown the importance of the FM stray fields in the overall magnetic
response of $Nb/Co$ superlattices. In that work we also hinted to the possibility that the SC response may modify the magnetic
state of the FM layers. In this letter we show that, indeed, the SC response modifies the magnetic state of the FM layers. The
system global electromagnetic response is determined by an interplay between the SC magnetization temperature dependence and the
FM magnetization time evolution.

\begin{figure}[hhhh]
\includegraphics[width=9cm]{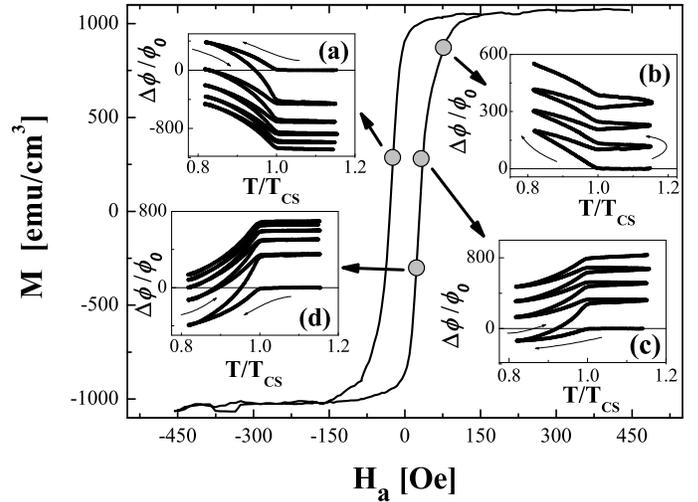}
\caption{\label{FigExpCompuesta}Main panel: Magnetization, $M$, as a function of applied field, $H_a$, for the
[$Nb$($44\,nm$)/$Co$($10\,nm$)]x19 superlattice at $T\,=\,7\,K\,>\,T_{CS}\,=\,6.2\,K$. Panels (a) through (d) show the
temperature, $T$, dependence of the superconducting magnetic flux response, $\Delta \phi$, in units of the superconducting flux
quantum, $\phi_o$, at different applied fields and ferromagnetic layers initial state, as indicated by gray dots and connecting
arrows in the main panel. Arrows in panels indicate the direction of the $T$ sweeps. Panel (a): $H_a\,=\,-22\,Oe$, +FC. Panel
(b): $H_a\,=\,75\,Oe$, -FC. Panel (c): $H_a\,=\,38\,Oe$, -FC. Panel (d): $H_a\,=\,22\,Oe$, -FC.}
\end{figure}

We present data on the temperature, $T$, dependence of the magnetic flux expulsion, $\Delta\phi$, directly proportional to the SC
magnetization, for $Nb$/$Co$ superlattices. Data is presented for a [$Nb$($44\,nm$)/$Co$($10\,nm$)]x19 and a
[$Nb$($44\,nm$)/$Co$($7.5\,nm$)]x19 superlattice. The results on these samples are representative of the measurements we
performed on a collection of $Nb/Co$ FM/SC superlattices. Since our experimental setup measures magnetic flux
variations,\cite{NuestroPRB} all $\Delta \phi$ values are measured with respect to the first data point, always above the
superconducting critical temperature, $T_{CS}$. Sample preparation, characterization method and measurement details are described
in reference \onlinecite{NuestroPRB}. For all flux expulsion data, the applied field, $H_a$, is parallel to the sample surface.
In the normal state, both superlattices present FM behavior with Curie temperatures, $T_C$, above $300\,K$. Flux expulsion in the
SC state was measured as a function of $T$  in field cooling experiments, for two different $Co$ layer´s initial FM states, -FC,
with the $Co$ layers initially saturated in the negative $H_a$ direction, and +FC, with the $Co$ layers initially saturated in
the positive $H_a$ direction. A detailed explanation of these measurement protocols is also included in reference
\onlinecite{NuestroPRB}.

Figure \ref{FigExpCompuesta} is a composite that summarizes the experimental results. In the main panel we show the FM hysteresis
loop of the $Co$ layers at $T\,=\,7\,K$, close but above $T_{CS}$ of the $Nb$ layers. In the superimposed panels, we show the $T$
dependence of $\Delta\phi$ for +FC and -FC measurements. For each initial state, several $T$ cycles were measured, each cycle
sweeping $T$ down from $7\,K$ to $5.5\,K$ and up to $7\,K$ again. The solid dots in the hysteresis curve connected with arrows to
the panels indicate the initial magnetic state for each experiment.

The data for the first $T$ down-sweep in each panel shows the behavior already discussed in our previous work.\cite{NuestroPRB}.
As discussed there, the SC response is proportional to the effective field, $H_{eff}$, originated by the superposition of the
applied field, $H_a$, and the $Co$ layers´ stray field, $H_s$.  For the -FC measurements (lower branch of the $Co$ hysteresis
loop), at low $H_a$ and negative $Co$ magnetization, the SC layers sense a positive $H_{eff}$ due to the $Co$´s $H_s$, see panel
(d). At higher $H_a$, the $Co$ magnetization becomes positive, $H_s$ becomes negative and larger than $H_a$, and the SC senses a
negative $H_{eff}$, see panel (b). Panel (c) shows an intermediate case, where the magnetization is already reversed, but $H_s$
is smaller than $H_a$ and the SC still senses a positive $H_{eff}$. Panel (a), +FC initial state (upper branch of the $Co$
hysteresis loop) is the mirror experiment from panel (d).

The novel feature observed in these data is present in the dependence of the $Co$ magnetization in the normal state with the
number of $T$ sweeps, i.e. cycles. This dependence is observed as a non-repeatability of the normal state magnetization value
after a cycle is completed. This behavior is not due to an experimental artifact related to an instrumental drift, since this
instrumental drift has been substracted from the data. A systematic behavior is observed in spite of the seemingly complex
dependence. The direction of the variation follows the sign of the \textit{applied} field, and is independent of the $Co$
magnetization direction, i.e. the \textit{stray} field, compare data in panel (b) and (d), for example. The difference between
the first and second normal state $\Delta \phi$ values in panel (d) of $360$ superconducting flux quantums is equivalent to a
change of $1.1\,emu\,cm^{-3}$ in the $Co$ layer´s magnetization, which shows that this effect is small but not negligible.

\begin{figure}[hhhh]
\includegraphics[width=8cm]{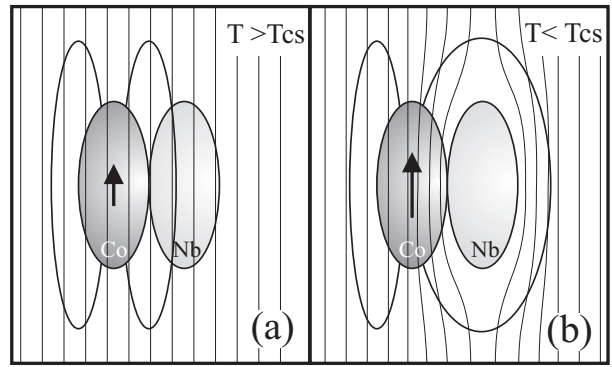}
\caption{\label{FigModelo}Schematics of the ``toy model'' behavior. Panel (a): at $T\,>\,T_{CS}$ the normal $Nb$ ellipsoid
experiences an effective field due to the applied field (straight lines) and the ferromagnetic $Co$ ellipsoid stray field (dipole
like lines). Panel (b): As $T$ is reduced below $T_{CS}$ the magnetic flux expulsion from the $Nb$ ellipsoid modifies the
effective field over the $Co$ ellipsoid.}
\end{figure}

In order to understand this behavior, we have constructed a simple ``toy model'' to qualitatively simulate the experimental data.
Although the model is very simple, a careful consideration of its hypothesis should be made to fully understand the implications
of the results. The first requirement is that an electromagnetic stray-field mediated interaction should exist between the FM and
the SC components. This is not achievable if the materials are modelled as nearly infinite slabs parallel to the applied field,
since the stray field of this geometry is negligible. Consequently, both materials are modelled as ellipsoids with one of the
principal axis parallel to the field. The ``toy model'' sample consists, then, of a FM and a SC ellipsoids, located side by side.
Figure \ref{FigModelo} depicts the main ideas of the model. Panel (a) shows the situation at $T\,>\,T_{CS}$ where the $H_{eff}$
sensed by the SC ellipsoid is composed by $H_a$ (straight lines) and the FM $H_s$ (dipolar lines arising from the FM ellipsoid).
When $T$ is reduced below $T_{CS}$, as depicted in panel (b), the flux expulsion from the SC ellipsoid modifies the $H_{eff}$
sensed by the FM ellipsoid and consequently its magnetization. The solution of the problem has now to be found in a
self-consistent way.

The ellipsoid shape or eccentricity, $\epsilon$, was selected as to maximize the stray field effects. That an optimum value
exists is clear from the fact that in the $\epsilon \rightarrow \infty$ ``needle'' limit, the stray fields approach zero due to
the negligible demagnetizing effects, and that in the $\epsilon \rightarrow 0$ ``disk'' limit, the stray fields also approach
zero since the ellipsoid is being magnetized along the shape anisotropy ``hard axis''. The optimal $\epsilon$ value actually
depends on the material´s magnetization, but since it is weakly dependent on it, a value of $10$ was found to maximize the stray
field effects in nearly all the $T$-$H_a$ range. Also, since an exact three dimensional spatial solution of this electromagnetic
problem is beyond the scope of this work, and would only obscure the results of the model, the spatial dependence of the stray
fields is neglected, and $H_s$ due to each ellipsoid is evaluated only at the center of the other ellipsoid.

The second ingredient in the model is a ``time'' dependence. This dependence cannot be ascribed to the superconducting material
since we have shown that no vortices are present in the $T-H_a$ range of these experiments. Consequently, it must be arising from
the creep in the FM material. Following this idea, the magnetization of the SC $Nb$ ellipsoid is modelled by a $T$ dependent,
time independent Meissner state. As a further simplification of the model the $T$ dependence is forced to follow that of a
parallel slab with a two fluid $T$ law.\cite{NuestroPRB} On the other hand, the FM $Co$ ellipsoid magnetization does not present
a $T$ dependence since its $T_C$ is much higher than the measurement range. It only shows a time dependence which must be
numerically simulated, as described in the next paragraph.

To simulate the $T$ sweeps at constant $H_a$, the self-consistent equilibrium state of the magnetized ellipsoids is solved at a
given $T$. After this, the magnetization change for the $Co$ ellipsoid, is reduced by a given percentage, and the magnetization
of the $Nb$ ellipsoid is recalculated for this, now fixed, value of the $Co$ ellipsoid´s magnetization, i.e. stray field. This
algorithm results in an effective exponential time dependence for the $Co$ magnetization. The sample´s magnetization, $M_T$, is
defined as the total magnetic moment divided by the total sample volume. In order to compare the results to the experiments, the
simulation data is presented as $\Delta\,M_T\,=\,M_T\,-\,M_o$, where $M_o$ is the value for the first simulated point, always at
$T\,>\,T_{CS}$.

\begin{figure}[hhhh]
\includegraphics[width=8cm]{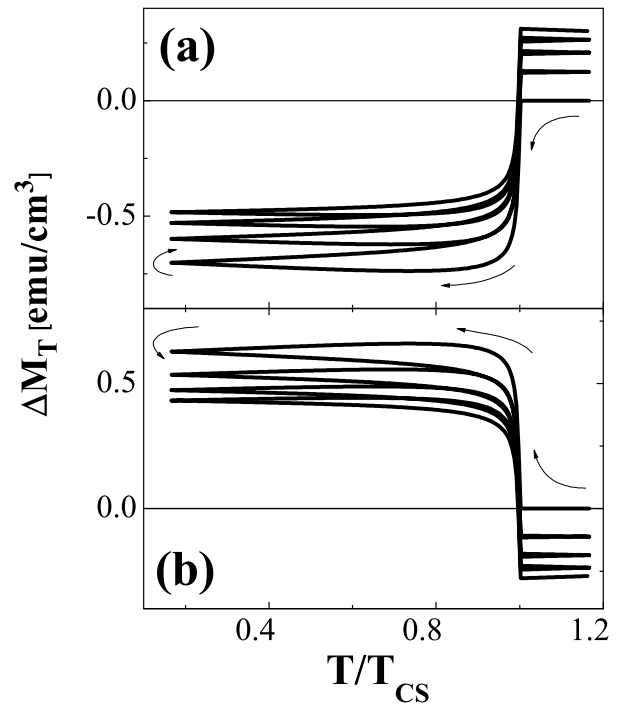}
\caption{\label{FigResModeloCompuesta}Model prediction for the temperature, $T$, dependence of the sample magnetization change,
$\Delta M_T\,=\,M_T(T)-M_o$, where $M_o$ is the magnetization for the initial simulated data point at $T\,>T_{CS}$. Panels (a)
and (b) show the results for two sets of parameters qualitatively equivalent to the experimental data in panels (c) and (a) of
figure \ref{FigExpCompuesta}. Panel (a): $H_a\,=\,22.85\,Oe$, $M_o\,=\,5.12\,emu\,cm^{-3}$. Panel (b): $H_a\,=\,-17.15\,Oe$,
$M_o\,=\,5.12\,emu\,cm^{-3}$}.
\end{figure}

Panels (a) and (b) in figure \ref{FigResModeloCompuesta} show the prediction of the model for situations similar to panels (c)
and (a) in figure \ref{FigExpCompuesta}, i.e. opposite direction of $H_a$ and same value of $M_o$. It is clear that the principal
features of the experimental data are qualitatively reproduced. First, there is a dependence of the normal state magnetization
with the number of cycles. This dependence follows the sign of the applied field and is not correlated to the magnetization
direction. Second, there is an irreversibility between cooling-down and warming-up sweeps. Third, a non-monotonic $T$ dependence
is observed for cooling-down sweeps.

\begin{figure}[hhhh]
\includegraphics[width=9cm]{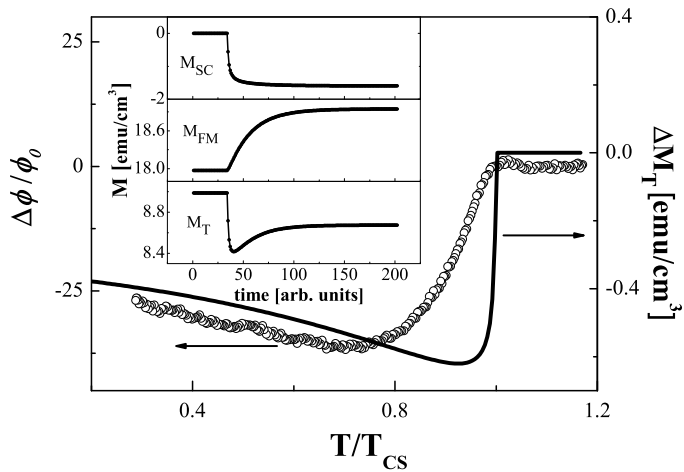}
\caption{\label{FigResModeloTiempo}Magnetic flux expulsion, $\Delta \phi$ for a [$Nb$($44\,nm$)/$Co$($7.5\,nm$)]x19 superlattice
with $T_{CS}\,=\,5.9\,K$, and model prediction, $\Delta M_T$, for $H_a\,=\,15\,Oe$ and $M_o\,=\,9\,emu\,cm^{-3}$. Inset shows the
dependence on simulated data point number, i.e.``time'', for the superconducting ellipsoid magnetization, $M_{SC}$, the
ferromagnetic ellipsoid magnetization, $M_{FM}$, and the sample´s total magnetization, $M_T$.}
\end{figure}

An interesting feature not actually observable in the data in fig.\ref{FigExpCompuesta}, but presented in reference
\onlinecite{NuestroPRB} is a non-monotonic T dependence that develops for applied fields near the coercive field of the $Co$
layers. The main panel in figure \ref{FigResModeloTiempo} shows a comparison between experimental and simulated data, where the
simulation parameters have been selected as to maximize this non-monotonic $T$ dependence. The origin of this behavior becomes
clear when examining separately the $Nb$ and $Co$ magnetization response in the simulated data. The inset shows the SC ellipsoid
magnetization, $M_{SC}$, the FM ellipsoid magnetization, $M_{FM}$, and the sample´s magnetization, $M_T$, as a function of
simulated data point number, i.e.``time'', while $T$ is swept down from above $T_{CS}$. The $T$ sweep is linear with this
``simulated time''. The time dependence of $M_{SC}$ is that arising from the $T$ sweep, given that the Meissner state does not
present an intrinsic time dependence. The time dependence of $M_{FM}$, on the other hand, has a twofold origin. First, the flux
expulsion in the SC originates an increase of local magnetic field in the FM material, as schematized in panel (b) of figure
\ref{FigModelo}. Second, the $M_{FM}$ presents an intrinsic time dependence in its response to the magnetic field changes. In
this light, the origin of the non-monotonic $T$ dependence becomes clear. As $T$ is swept down from above $T_{CS}$, the $T$
dependence of the SC ellipsoid magnetization produces a flux expulsion in the sample. This originates a field increase in the FM
material increasing its $M_{FM}$. At lower temperatures, the $T$ dependence of the SC material is relatively weak, and the time
dependence of the FM material emerges as a ``paramagnetic'' like signal, resembling a paramagnetic Meissner
effect.\cite{ParMeissnerWolleben,JacoboParMeissner,DiscoNbParMeissEff}.

The results and the toy model presented here clarify the response of SC/FM hybrid structures and, at the same time, raise an
interesting question. We have demonstrated that the electrodynamic response of these hybrid systems involves a combination of two
separate phenomena. In the first place, the diamagnetic response of the SC layers expels the magnetic flux into the FM layers. As
a consequence, the FM material responds with a time dependence, clearly in the direction of the applied field. Both materials
affect each other with their respective stray fields. In this process, the magnetic domain structure of the FM seems to play an
important role, since the stray fields of an infinite slab are negligible. Clearly, in order to observe stray field effects, a
non slab geometry has to be present in the samples. In this picture, an interesting point arises. Given that the response of the
hybrid material is affected, and in some $T$ and $H$ range, dominated, by the intrinsic time dependence, the effects described
here may be important if the device operation is based on magnetization \textit{changes} and designed to work at frequencies
similar to the creep of the FM.

In summary, we have demonstrated that the electrodynamic response of SC/FM hybrid materials is determined by an interplay between
the temperature dependence of the SC magnetization, the time dependence of the FM magnetization, and the effective interaction
between them mediated by the stray fields.

Work partially supported by ANPCyT PICT2003-03-13511, ANPCyT PICT2003-03-13297 and Fundación Antorchas. CM acknowledges financial
support from and JG is a member of CONICET, Argentina.

\end{document}